\documentclass[reprint,nofootinbib,amsmath,amssymb,showkeys,aps]{revtex4-1}
\usepackage{ulem}
\usepackage{graphicx}
\usepackage{graphicx}
\usepackage{epstopdf}
\usepackage{dcolumn}
\usepackage{bm}
\usepackage{comment}
\usepackage[utf8]{inputenc}
\usepackage{mciteplus}
\usepackage{color}
\usepackage{enumitem}
\usepackage{subfigure}
\usepackage{braket}
\usepackage{enumitem}
\usepackage[colorlinks]{hyperref}
\usepackage[color=lightgray]{todonotes}
\hypersetup{linkcolor=blue,filecolor=blue,urlcolor=blue,citecolor=blue}
\newcommand{\lsim}{\raisebox{-0.13cm}{~\shortstack{$<$ \\begin{equation}-0.07cm]  $\sim$}}~}
\newcommand{\gsim}{\raisebox{-0.13cm}{~\shortstack{$>$ \\begin{equation}-0.07cm] $\sim$}}~}

\begin{document}
 \vspace*{-5mm}
\begin{flushright}
HIP-2019-18/TH
\end{flushright}
\vspace{.3cm}
\thispagestyle{empty}
\title{\boldmath Primordial dark matter halos from fifth forces}

\author{Stefano Savastano$^{1,3}$}
\email{stefano.savastano@studio.unibo.it }
\author{Luca Amendola$^1$}
	\email{l.amendola@thphys.uni-heidelberg.de}
\author{Javier Rubio$^{2}$}
 	\email{javier.rubio@helsinki.fi}
\author{Christof Wetterich$^1$}
\email{c.wetterich@thphys.uni-heidelberg.de}

\affiliation{$^1$ Institut f{\"u}r Theoretische Physik, Ruprecht-Karls-Universit{\"a}t Heidelberg,  Philosophenweg 16, 69120 Heidelberg, Germany }
\affiliation{$^2$ Department of Physics and Helsinki Institute of Physics, \\ 
PL 64, FI-00014 University of Helsinki, Finland}
\affiliation{$^3$ Dipartimento di Fisica e Astronomia, Università di Bologna, \\ Via Irnerio 46, 40126 Bologna, Italy}

\begin{abstract}
We argue that primordial dark matter halos could be generated during radiation domination by long-range attractive forces stronger than gravity. In this paper, we derive the conditions under which these structures could dominate the dark matter content of the Universe while passing microlensing constraints and cosmic microwave background energy injection bounds. The dark matter particles would be clumped in objects in the solar mass range with typical sizes of the order of the solar system. Consequences for direct dark matter searches are important. 
\end{abstract}

\maketitle

\section{Introduction}

The formation of bound objects in the standard cosmological $\Lambda$CDM scenario is restricted to small redshifts. This result is based on i) gravity being the dominant attractive force for the clumping of matter ii) the assumption of a nearly scale-invariant spectrum of primordial density perturbations at all scales. These two assumptions entail the absence of significant structure formation prior to matter-radiation equality. None of these conditions must be necessarily fulfilled in alternative cosmologies. Strong deviations from scale invariance leading to the formation of ultracompact minihalos (UCMHs) \cite{Berezinsky:2003vn,Bringmann:2011ut}, axion miniclusters \cite{Hogan:1988mp,Fairbairn:2017sil} or primordial black holes (PBHs) \cite{1966AZh....43..758Z,Hawking:1971ei,Carr:1975qj,1975Natur.253..251C,Carr:2016drx,Khlopov:2008qy} are expected to appear, for instance, in scenarios displaying nontrivial features along the inflationary trajectory \cite{GarciaBellido:1996qt,Yokoyama:1995ex,Drees:2011hb,Kawasaki:2016pql,Frampton:2010sw,Garcia-Bellido:2017mdw,Kannike:2017bxn,Germani:2017bcs,Motohashi:2017kbs,Ballesteros:2017fsr}. Alternatively, compact objects could be generated by the action of an additional attractive force stronger than gravity, able to enhance the growth of perturbations during matter or radiation domination. Light scalars are a natural possibility for mediating such a force. A realization of this scenario was recently advocated in Ref.~\cite{Amendola:2017xhl} (see also Ref.~\cite{Bonometto:2018dmx}). The main ingredient of the proposal was the existence of a long-range interaction mediating between particles in a beyond the Standard Model sector and leading eventually to the formation of primordial black holes. In this paper, we focus on an alternative outcome of the scenario:  the formation of primordial dark matter halos (PDMHs).  

We consider a specific implementation of the above fifth force framework based on a light scalar field---potentially, but not necessarily, identified with a dynamical dark energy component---and a beyond the Standard Model fermion playing the role of cold dark matter. The two species are assumed to be subdominant with respect to the Standard Model component during the relevant cosmological epochs, i.e. before and during PDMH formation. The fermions couple to the scalar field, which mediates an attraction that can be stronger than gravity, as typically happens in variable gravity scenarios \cite{Wetterich:1987fk,Wetterich:2013jsa,Rubio:2017gty}. For sufficiently strong coupling, the system approaches an attractor solution during radiation domination where the subdominant scalar and fermion components track the background energy density, such that the cosmological fractions of the three species remain constant \cite{Wetterich:1994bg,Amendola:1999er,TocchiniValentini:2001ty,Amendola:2001rc,Bonometto:2012qz,Amendola:2017xhl}. This solution has a strong impact on the evolution of fermionic density perturbations, which start to grow under the action of the fifth force and eventually lead to the formation of virialized halos with a mass only depending on the strength of the fermion-scalar coupling. 
 
 The mass of the dark matter fermion decreases as the scalar field changes with time. As an example, it may change from $1$ MeV to $0.1$ keV between the onset of the scaling regime and virialization. This corresponds to a scalar mediated attraction 100 times stronger than gravity and a final mass of the bound objects constrained by observations to lie between $10^{-8}$ and $10^4$ $M_\odot$.

This paper is organized as follows. The main ingredients of the model are reviewed in Sec. \ref{sec:5thforce} where, upon discussing the background evolution, we extend the treatment of fluctuations in Ref.~\cite{Amendola:2017xhl} to the nonlinear regime. The conditions leading to the formation of primordial dark matter halos are discussed in Sec. \ref{sec:properties}, where we present analytical estimates for the virialization radius, the mass-radius relation and the properties of the constituent particles. The comparison of the fifth force created structures with observations is performed in Sec. \ref{sec:constraints}. Finally Sec. \ref{sec:conclusions} contains our conclusions.

\section{Fifth force interactions}\label{sec:5thforce}

\noindent We consider a minimal extension of the Standard Model with Lagrangian density
\begin{equation}\label{eq:action}
\frac{\cal L}{\sqrt{-g}}=\frac{M_{P}^{2}}{2}R+{\cal L}_{\rm R}+{\cal L(\phi)}+{\cal L} (\phi,\psi)\,.
\end{equation} 
Here $M_{P}=(8\pi G)^{-1/2}=2.435\times 10^{18} \,{\rm GeV}$ is the reduced Planck  mass, $R$ is the Ricci scalar and ${\cal L}_{\rm R}$ denotes a Standard Model radiation component that we assume to dominate the Universe at early times.  The term
\begin{eqnarray}\label{eq:action2}
&&{\cal L(\phi)}=-\frac{1}{2}\partial^{\mu}\phi\partial_{\mu}\phi-V(\phi)
\end{eqnarray}
stands for the Lagrangian density of a canonically normalized scalar field $\phi$. This beyond the Standard Model component is taken to be  interacting with a fermion field $\psi$ via a field-dependent mass term $m_\psi(\phi)$, 
\begin{eqnarray}\label{eq:action3}
&&{\cal L(\phi,\psi)}=i\bar{\psi}\left(\gamma^{\mu}\nabla_{\mu}-m_\psi(\phi)\right)\psi\,.
\end{eqnarray}
The interaction strength is given by an effective coupling
\begin{equation}
\beta(\phi)\equiv -M_{P}\frac{\partial\ln m_\psi(\phi)}{\partial\phi}\,,
\label{betadef1}
\end{equation}
measuring the change of the fermion mass with the scalar field $\phi$.
For $\vert\beta\vert \approx 1$ this coupling mediates an attraction of gravitational strength. The typical values of $|\beta|$ considered in this paper will be, however, larger than unity, leading therefore to a pull stronger than gravity and an additional power injection mechanism in this sector. In particular, we will consider a range $3\lesssim \beta \lesssim 30$ in order to pass several observational constraints that will be  discussed below.~\footnote{For concreteness, we assume $\beta>0$, although the scenario remains qualitatively valid for negative and large values of $\beta$ as well.}

The effective coupling \eqref{eq:action3} generates an energy-momentum transfer among the scalar and fermion components, namely
\begin{eqnarray}
\nabla_{\nu}T_{(\phi)}^{\mu\nu}&=&  \frac{\beta(\phi)}{M_P}T_{(\psi)}\partial^{\mu}\phi\,,\label{eq:continuity1}\\ 
\nabla_{\nu}T_{(\psi)}^{\mu\nu}&=&  -\frac{\beta(\phi)}{M_P}T_{(\psi)}\partial^{\mu}\phi \label{eq:continuity2}\,,
\end{eqnarray}
with $T_{(\psi)}=T_{(\psi)}^{\mu\nu}g_{\mu\nu}$ the trace of the $\psi$-field energy momentum tensor. This type of scenario has been extensively studied in the literature \cite{Wetterich:1994bg,Amendola:1999er,Amendola:2003wa,Fardon:2003eh,Brookfield:2005bz,Koivisto:2005nr,Amendola:2006dg,Wetterich:2007kr,Amendola:2007yx,Boehmer:2008av,Baldi:2008ay,Mota:2008nj,Pettorino:2009vn,Wintergerst:2009fh,Baldi:2010vv,Ayaita:2011ay,Nunes:2011mw,Ayaita:2012xm,Bonometto:2012qz,Bonometto:2013eva,Ayaita:2014una,Fuhrer:2015xya,Bonometto:2015mya,Maccio:2015iya,Gleyzes:2015pma,Casas:2016duf,Bonometto:2017rdu,Bonometto:2017lhg,Amendola:2017xhl,Bonometto:2018dmx}.
In distinction to growing neutrino quintessence models \cite{Amendola:2007yx}, the mass of the dark matter fermion in the scenario at hand is in the MeV range and is therefore much larger than neutrino masses. This leads to distinct characteristic length scales and time distances. Different choices of $\beta(\phi)$ correspond to different realizations. A simple  possibility is to consider an effective coupling $\beta (\phi)=-g M_P/(m_0+g\phi)$
following from a renormalizable Yukawa interaction $m_\psi(\phi)\bar\psi \psi=m_0\bar \psi\psi +g \phi \bar \psi\psi$,  with $m_0$ a mass parameter and $g$ a dimensionless coupling. A value of $|\beta|$ substantially larger than unity follows even for small $g$ if the fermionic mass term $m_0$ is sufficiently below $M_P$.
Alternatively, one could consider a setup involving a constant $\beta$ coupling. This describes dilatoniclike interactions 
\begin{equation}
    m_\psi(\phi)\bar\psi\psi=m_0 \exp\left(-\beta \phi/M_P\right)\bar\psi\psi
    \label{eq:mphi}
\end{equation}
as those naturally appearing in scalar-tensor theories when written in the Einstein frame \cite{Wetterich:1987fk,Wetterich:2013jsa,Rubio:2017gty}. For the sake of simplicity, we will restrict ourselves to the latest possibility, understanding it as an approximation of the real dynamics for the relevant temporal scales and in the absence of significant backreaction effects.

\subsection{Background evolution}\label{sec:background}  

\begin{figure}
    \centering
    \hspace{-0.3cm}\includegraphics[scale=0.7]{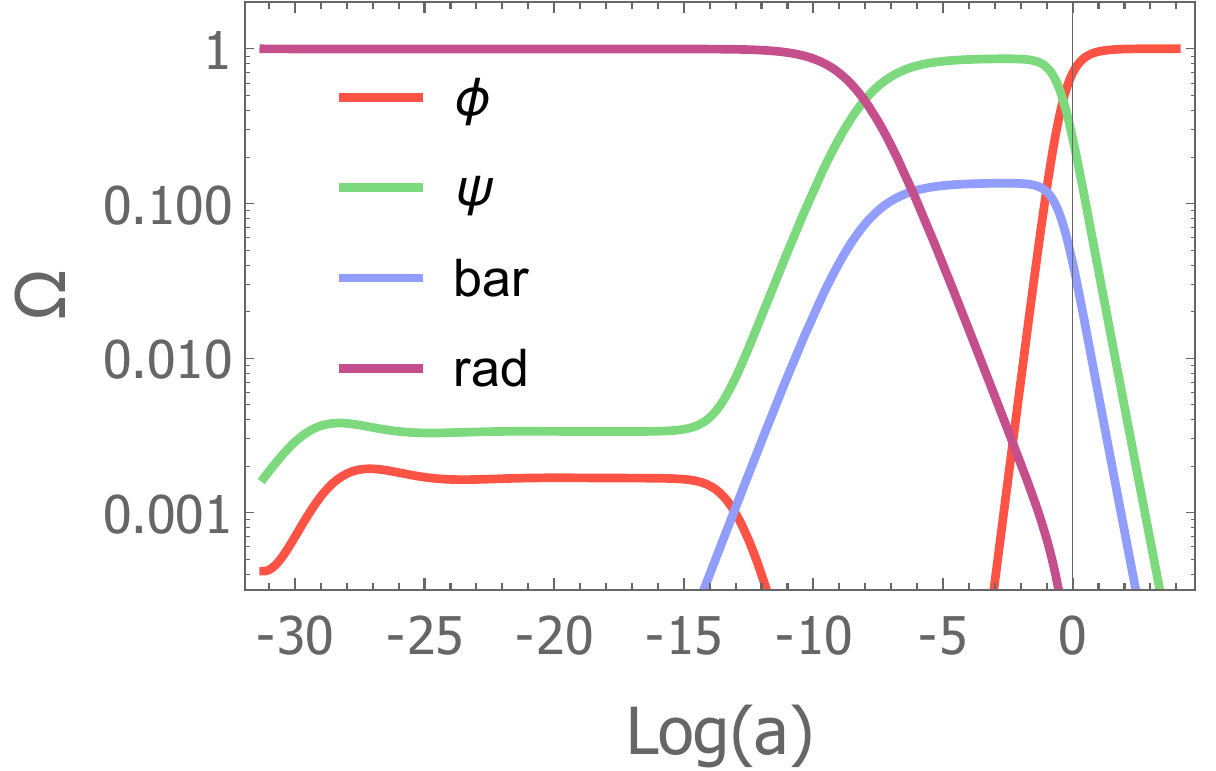}
    \caption{Evolution of the different density fractions for ${\beta=10}$ and an exponential potential for the scalar field $\phi$. The coupling $\beta$ is switched off at virialization (cf.~Sect.~\ref{sec:properties}), here taken to occur at a redshift $z=10^6$. After a transition phase, the energy densities of $\phi$ and $\psi$ components set on the scaling solution during the radiation-dominated epoch. After the scaling regime the evolution is very close to standard $\Lambda$CDM.}
    \label{fig:beta10}
\end{figure}

Assuming a flat Friedmann-Lema\^itre-Robertson-Walker universe and a perfect fluid description, the background evolution equations following from the expressions \eqref{eq:continuity1} and \eqref{eq:continuity2} can be written as
\begin{eqnarray}
 && \dot{\rho}_{\phi}+3H(\rho_{\phi}+p_{\phi})=\,\frac{\beta}{M_{P}}\left(\rho_{{\rm \psi}}-3p_\psi\right)\dot{\phi}\,,\label{eq3a}\\
 && \dot{\rho}_{{\rm \psi}}+3H\left(\rho_{{\rm \psi}}+p_\psi\right)=-\,\,\frac{\beta}{M_{P}}\left(\rho_{{\rm \psi}}-3p_\psi\right)\dot{\phi}\,,\label{eq3}
\end{eqnarray}
with $H$ the Hubble rate and $\rho_i$ and $p_i$ the average energy density and pressure of the $i=\phi,\psi$ components. We consider a scenario where the radiation fluid is the dominant energy component during the period of formation of dark matter halos. Both the heavy fermion and the scalar field constitute therefore a subleading fraction of the total energy density of the Universe and will adapt their evolution to the dominant radiation counterpart.
The interaction term at the right-hand side of Eqs.~\eqref{eq3a} and \eqref{eq3} is active  whenever the $\psi$ particles are nonrelativistic (i.e. for $T_{(\psi)}\neq 0$ or $\rho_\psi\neq 3 p_\psi$). 
In this limit and for $\beta\gg1$, the model admits an attractor solution where the scalar and fermion energy densities \textit{track} the background radiation component (see Fig.~\ref{fig:beta10}). During this regime  we have \cite{Wetterich:1994bg,Amendola:1999er,TocchiniValentini:2001ty,Amendola:2001rc,Bonometto:2012qz,Amendola:2017xhl}
 \begin{equation}\label{eq:bphi}
\phi'=M_P/\beta\,,  
\end{equation}
 with the prime denoting derivatives with respect to the number of $e$-folds $dN\equiv H\,dt$ and
\begin{equation}\label{eq:phaseq-1}
\Omega_{\psi}=\frac{1}{3\beta^{2}}\,,\hspace{8mm}
\Omega_{\phi}=\frac{1}{6\beta^{2}}\,,\hspace{8mm}
  \Omega_{R}=1-\frac{1}{2\beta^{2}}\,.
\end{equation}
Here $\Omega_{i}\equiv \rho_{i}/(3M_{P}^{2}H^{2})$ stands for the energy density parameters for the $i=R,\phi,\psi$ components  and $\rho_i\sim a^{-4}$. During the scaling solution the fermion mass decreases according to
\begin{equation}\label{eq:fermass}
    m_\psi'=\frac{d \,m_\psi}{d\phi}\phi'=-\beta \frac{m_\psi}{M_P}\phi'=-m_\psi \hspace{3mm}\Longrightarrow  \hspace{3mm} m_\psi\sim a^{-1}
\end{equation}
and independently of $\beta$.

 Depending on the initial conditions following the end of inflation, the fixed point \eqref{eq:phaseq-1} could be reached immediately after this era or at later times. We denote by $a_{\rm in}$ the scale factor at the time the scaling solution is reached. We will discuss how ``initial values" of the density contrast at $a_{\rm in}$ will grow and form extended objects.
 The time at which the scaling solution is reached will be an important parameter for setting the characteristic scales of our scenario. As shown in detail in Sec. \ref{sec:properties}, if the PDMHs constitute the entire dark matter component, a typical redshift at which the scaling solution has to set in for a fiducial coupling $\beta=4$ is 
 \begin{equation}
z_{\rm in}\approx 3\cdot 10^8\,.
 \end{equation}
 Assuming thermal equilibrium, the masses of the $\psi$-particles at this time are of order $m_\psi(z_{\rm in})\sim {\cal O}({\rm MeV})$ or larger. If this hypothesis is dropped, the estimate of the mass scale becomes more complicated. Given a universal reheating production at the end of inflation, one would expect an initial momentum distribution in the $\psi$-sector similar to that of photons. Then, even if not in thermal contact, the two species could have maintained a similar temperature, except for the subsequent increase of the photon entropy due to pair annihilation. In this case, the mass of the $\psi$-particles at $z_{\rm in}$ must exceed the photon temperature since the $\psi$-particles need to be non relativistic for the existence of the scaling solution. Our estimate for the lower bound on the mass remains valid as an order of magnitude.
 
 Once the scaling solution is reached, it can extend up to matter-radiation equality. The main restriction to this possibility is associated with big bang nucleosynthesis.  In particular, the presence of the additional relativistic components modifies the expansion rate of the Universe as compared to the standard hot big bang theory and with it the relative abundance of light elements. The tight constraints on these quantities translate into an upper bound on the density parameters, $\Omega_\phi\vert_{\rm BBN}+\Omega_\psi\vert_{\rm BBN} < 0.045$  \cite{Bean:2001wt}, roughly corresponding to a mild restriction $\beta\gtrsim 3$. This constraint can be evaded if the $\psi$-particles become non relativistic only after big bang nucleosynthesis\footnote{This could happen for instance, if they were in thermal equilibrium and had a mass much smaller than 0.1 MeV.} since the corresponding density parameter would be then the very small one inherited from the primordial abundance and not the one dictated by the attractor solution \eqref{eq:phaseq-1}. For simplicity, however, we will conservatively assume the above restriction on $\beta$. 

\subsection{Growth of fluctuations}\label{sec:growth}

In a standard gravitational context, the density contrast evolution can be inferred from the Navier-Stokes equations. For coupled cosmologies, these equations extend to \cite{Amendola:2003wa}
\begin{eqnarray}
\delta_{\psi} ' & = & -\nabla_i(1+\delta_\psi)v^i_{\psi}  \,,\label{eq:deltapsi} \\[0.3cm]
v^{i'}_{\psi} & = & -\left(1+\frac{\mathcal{H}'}{\mathcal{H}}-\sqrt{6}\beta\right)v^i_{\psi}+ \label{eq:vel}\\[0.1cm] \nonumber
&&\hspace{0.5cm}-v^j_{\psi}\nabla_j v^i_{\psi}-\mathcal{H}^{-2}\nabla^i \hat{\Phi}\,, \\[0.3cm]
\Delta\hat{\Phi}& = &\frac{3}{2}\mathcal{H}^2(Y\delta_\psi\Omega_\psi+\Omega_{R}\delta_{R})\,,\label{eq:modposs} 
\end{eqnarray}
where we have defined  the density contrasts $\delta_{R,\psi}$ for radiation and $\psi$, respectively, and a velocity field
\begin{equation}
    v^{i}_\psi=\frac{{x}^{i\,'}}{2a\mathcal{H}}\,.
\end{equation}
Here, $x_i$ are the co-moving coordinates, ${\cal H}\equiv a H$ is the conformal Hubble rate and the bar refers to the background value. 
The \textit{modified} Newtonian potential $\hat{\Phi}\equiv \Phi-\sqrt{6}\delta\phi$ is sourced by the  $\psi$-field fluctuations  via the \textit{modified} Poisson equation \eqref{eq:modposs}, with 
\begin{equation}\label{eq:Ydef}
Y\equiv 1+2\beta^2
\end{equation}
an effective coupling encoding the combined strength of the fifth force and gravity. This force equals the gravitational pull for $\beta=1/\sqrt{2}$ and becomes significantly stronger than it for $\beta\gg 1/\sqrt{2}$. In Eqs.~\eqref{eq:deltapsi}-\eqref{eq:modposs} we have  neglected the small contribution of $\phi$ perturbations since they do not experience a significant growth due to their unit speed of sound.

\begin{figure}
\begin{center}
\includegraphics[scale=0.80]{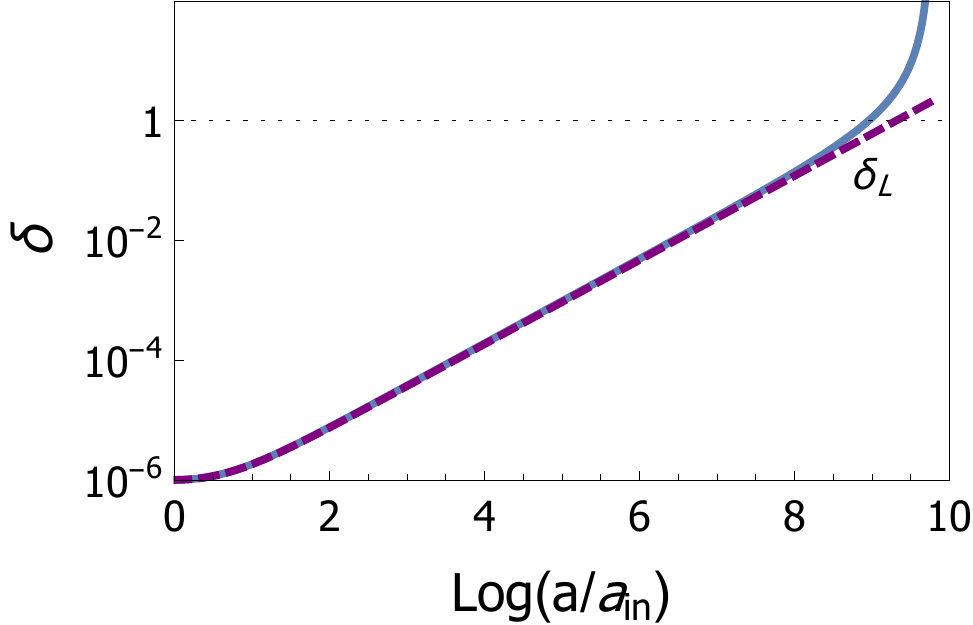}
\caption{Comparison between the growth of overdensities following from Eq.~\eqref{eq:delta_psi} and its linearized counterpart. The dashed-horizontal line $\delta=1$ is added for reference. Here $a_{\rm in}$ denotes the onset of the scaling solution.}
\label{fig:SolnD}
\end{center}
\end{figure}

\textit{In the absence of shear or rotational components in the initial velocity field}, the set of Eqs. \eqref{eq:deltapsi}-\eqref{eq:modposs} can be compacted in a single differential equation describing the non linear growth of matter density fluctuations,
\begin{eqnarray}\label{PertEqn}
\delta_\psi''&+&\left(1+\frac{\mathcal{H}'}{\mathcal{H}}-\frac{\beta\phi'}{M_p}\right)\delta_\psi' \\ &-&\frac{3}{2}(Y\delta_\psi\Omega_\psi+\Omega_{R}\delta_{R})(1+\delta_\psi)-\frac{4}{3}\frac{\delta_\psi'^2}{(1+\delta_\psi)}=0\,. \nonumber
\end{eqnarray}
During the scaling regime \eqref{eq:phaseq-1} the background evolution of the Universe is essentially dominated by the radiation component and we can safely approximate ${\cal H}'\simeq -{\cal H}$.\footnote{Accounting for the variation in the number of relativistic degrees of freedom $g(a)$ as the Universe expands has a minimal impact in this result. Indeed,  denoting $\gamma(a)=(g/g_{\rm eq})^{-1/3}$ with  $g_{\rm eq}\approx 3.36$, one has ${\cal H}'\simeq -{\cal H}(1-\gamma'/(2\gamma))$. In the temperature range $T=100$ GeV to $T=0.1$ MeV, the correction is smaller than 0.12 and can be safely neglected.} Taking this into account together with Eqs.~\eqref{eq:bphi} and \eqref{eq:phaseq-1},   Eq.~\eqref{PertEqn}, for large $|\beta|$, becomes independent of $\beta$,
\begin{equation}\label{eq:delta_psi}
\delta_\psi''-\delta_\psi'-(1+\delta_\psi)\delta_\psi-\frac{4}{3}\frac{\delta_\psi'^2}{(1+\delta_\psi)}=0 \,,
\end{equation}
 where we have neglected a small $\Omega_R\delta_R$ contribution. In Fig.~\ref{fig:SolnD} we show the numerical solution of Eq.~\eqref{eq:delta_psi} for $\delta_\psi$ as a function of the number of e-folds $N=\log(a/a_{\rm in})$.

At early times, the perturbations in the $\psi$ fluid are small and the linearized version of Eq.~\eqref{eq:delta_psi} admits a solution\footnote{As a funny coincidence, we note that $p$ equals the golden ratio $\varphi$.
The general solution for any $\beta$ is $p=(1\pm\sqrt{5 + 2\beta^{-2}})/2.$ We disregard the decaying mode. } \cite{Amendola:1999er,TocchiniValentini:2001ty,Amendola:2001rc,Bonometto:2012qz,Amendola:2017xhl}
\begin{equation}\label{eq:pertgrowth}
\delta_\psi=\delta_{\psi, \rm in}\left(\frac{a}{a_{\rm in}}\right)^{p}\,,\hspace{10mm} p=\frac{1}{2}\left(1 + \sqrt{5}\right)\approx 1.62\,,
\end{equation}
with $a_{\rm in}\equiv a(t_{\rm in})$ the scale  factor at the onset of the scaling regime.
The growth of initial inhomogeneities following from the sizable exponent $p$ brings them rapidly into a nonlinear regime. The precise onset of nonlinearities depends on the initial value $\delta_{\psi,\rm in}$, which should be \textit{a priori} determined by requiring compatibility with inflation. Some assumption about the \textit{full} initial power spectrum $\delta_{\psi,\rm in}$ is needed. In particular, the temperature fluctuations in the Cosmic Microwave Background (CMB) allow one to reconstruct the primordial power spectrum only at scales below the present horizon size and above a fraction of the sound horizon at recombination, namely $10-10^4$ Mpc. Although this limited range can be extended down to $ \sim 10^{-4} \,{\rm Mpc}$ by other measurements of the Lyman-$\alpha$ forest, weak gravitational lensing probes and spectral distortions \cite{Tegmark:2002cy,McDonald:2004eu,Chluba:2019kpb}, the amplitude and scale dependence of the primordial power spectrum is essentially unconstrained at the very small scales $10^{-13}\,{\rm Mpc}$ we will be interested in.
Contrary to standard PBHs and UCMHs formation scenarios, which rely on the assumption of a boosted primordial power spectrum at the scales of interest \cite{Berezinsky:2003vn,Bringmann:2011ut,1966AZh....43..758Z,Carr:2016drx}, we will adopt here a rather conservative point of view and assume the spectrum of $\psi$ perturbations to be commensurable with that of curvature perturbations at CMB scales. In the following sections, we explore the predictions of our model for different initial conditions, focusing for concreteness on a confidence interval $10^{-8} \leq\delta_{\psi,{\rm in}}\leq 10^{-4}$ at solar mass scales ${k_{\odot}\simeq10^{13}\,\textrm{Mpc}^{-1}}$ (cf. Sec. \ref{sec:massradius}).

\section{Primordial dark matter halos}\label{sec:properties}

 The evolution presented in the previous section should be understood just as an approximation of the real dynamics. On the one hand, the initial velocity perturbations in the $\psi$ fluid are expected to modify the simplistic spherical collapse and to favor the formation of virialized dark matter halos. On the other hand, the raising of the density within the collapsing regions is expected to trigger the screening of the fifth force.
 We will assume this to happen at some time between the onset of the scaling regime and virialization, so the created PDMHs stop growing and behave just as an ordinary dark matter fluid from thereon. There may be, however, some residual interaction of the dark sector with the scalar field, resulting in an effective coupling strength $\beta_{\rm eff}$ much smaller than $\beta$ \cite{Ayaita:2011ay,Nunes:2011mw,Ayaita:2012xm}. 
 
As we will discuss in detail in Sec. \ref{sec:screening}, we assume here that PDMHs are effectively screened after virialization ($a>a_{\rm V}$) and behave as non relativistic matter from there on, i.e. $\rho_{\rm PDMH}\sim a^{-3}$. Furthermore, we assume that the system evolution at virialization can be still approximated by the scaling solution (\ref{eq:phaseq-1}).
 During radiation domination, the PDMHs density parameter $\Omega_{\rm PDMH}(a)$ will consequently increase with time up to attaining a value 
 \begin{equation}\label{Omegah}
    \Omega_{\rm PDMH}(a_{\rm eq})=\Omega_{\rm PDMH}(a_{\rm V})\frac{a_{\rm eq}}{a_{\rm V}}
    =\frac{1}{3\beta^2}\frac{a_{\rm eq}}{a_{\rm V}}\,
\end{equation}
at matter-radiation equality. Here,  $a_{\rm eq}$ is the scale factor at equality and $a_{\rm V}$ denotes its value at virialization. In order to avoid the overclosure of the Universe, we will require $\Omega_{\rm PDMH}(a_{\rm eq})\leq 1/2$. While an additional dark matter component is generically needed if this inequality is not saturated, the created PDMHs can constitute the whole dark matter component in the Universe in the limiting case $\Omega_{\rm PDMH}(a_{\rm eq})=1/2$. In what follows, we will focus on this minimalistic possibility. In this case, the epoch of virialization is fixed by the condition 
\begin{equation}\label{eq:av3}
    \log\frac{a_{eq}}{a_{V}}=\log \frac{3\beta^2}{2}\approx 3\,,
\end{equation}
where for the last equality we have chosen a fiducial coupling $\beta=4$.

\subsection{Virialization}\label{sec:virialization}

The condition $\Omega_{\rm PDMH}(a_{\rm eq})=1/2$ in Eq.~\eqref{Omegah} translates into a consistency relation 
\begin{equation}\label{aVaeq}
\frac{a_{\rm V}}{a_{\rm eq}}=\frac{2}{3\beta^{2}}\,,    
\end{equation} 
meaning that virialization has to happen  well within radiation domination if $\beta\gg \sqrt{2/3}$. The precise onset of virial equilibrium is determined by the condition
\begin{equation}\label{vir_eqn}
 2K+U=0\,,   
\end{equation}
where $K$ and $U$ are the kinetic and the potential energies of the $\psi$-particle systems described as spherical overdensities. 

Following Ref.~\cite{Bonometto:2018dmx}, and in accordance with Birkhoff’s theorem, the potential energy experienced by the collapsing shell can be regarded as the sum of different contributions: first, the one sourced by the fermions and the scalar field on the overdense spherical region, which can be itself split into a background component and a perturbation component coupled also to the fifth force; and second, the potential energy sourced by the other background fluids contained in the shell. Accordingly, the potential experienced by a spherical overdensity becomes
\begin{equation}
    \frac{U(R)}{M}=-\frac{3}{5}G\frac{[\bar{M}+Y\delta M]}{R}-\frac{4\pi}{5}G(\rho_r+\rho_\phi)\,,
\end{equation}
or equivalently 
\begin{equation}\label{eq:UofY}
    \frac{U(R)}{M}=-\frac{3}{5}Y G\frac{\delta M}{R}-\frac{4\pi}{5}G\rho_{cr}R^2\,,
\end{equation}
with $\rho_{\rm cr}=3M_P^2 H^2$ the critical energy density and 
\begin{equation}
    \delta M=M-\bar{M}=\frac{4\pi}{3}\rho_{\rm cr}\Omega_\psi\delta_\psi R^3\,,
\end{equation}  
the difference between the overall shell mass $M$ and the background contribution $\bar M$. Combining Eq.~\eqref{eq:UofY} with the kinetic energy of the $\psi$-particles enclosed by the shell,
\begin{equation}
    K=\frac{3}{10}M\,\dot{R}^2=\frac{3}{10}M\,e^{-4N}R'^2\,,
\end{equation}
we can recast the virialization condition \eqref{vir_eqn} as
\begin{equation}  \label{eqn:eqnvir}
     2\,R'^2-\left[Y\Omega_{\psi}\delta_\psi(R,N)+1\right]R^2=0\,,
\end{equation}
with $\delta_\psi(R,N)$ the density contrast. 

The relation $\delta_\psi(R,N)$ in Eq.~\eqref{eqn:eqnvir} can be obtained by tracing the evolution of an initial spherical shell of radius $R_0$ enclosing a number of particles ${N_\psi=n_{\psi,{\rm 0}} R_{\rm 0}^3}$, with $n_{\psi,{\rm 0}}$ the initial particle density. Taking into account the scaling $n_\psi\sim a^{-3}$ and requiring the conservation of the number of particles within the shell,
\begin{equation}
n_{\psi} R^3=n_{\psi,{\rm 0}}R^3_{\rm 0}\,,
\end{equation}
we obtain
\begin{equation}\label{eq:Revol}
1+\delta_{\psi}(R,N)=\left(1+\delta_{\psi,\rm 0}\right)\left(\frac{R_{0}}{R}\right)^3 e^{3N}\,,\vspace{0.3cm}
\end{equation}
with ${N=\ln(a/a_{\rm in})}$ the number of e-folds of collapse. 

\subsection{Mass-radius relation} \label{sec:massradius} 

The energy density of the collapsing $\psi$ particles can be thought of as the sum of the contributions of $ n_\psi$ particles with field-dependent mass $m_\psi$, i.e. $\rho_\psi=n_\psi m_\psi$. Considering the scaling $n_\psi\propto a^{-3}$, together with the relativistic behavior of the $\psi$-field energy density during the tracking regime, $\rho_\psi\sim a^{-4}$, we get a temporal evolution
\begin{equation}\label{eq:massin}
m_\psi\propto a^{-1}\,,   
\end{equation}
in accordance with Eq.~\eqref{eq:fermass}.
This microscopical behavior translates into an effective change of the mass 
\begin{equation}\label{M0R}
    M_0\equiv \frac{4\pi}{3}\bar \rho_\psi R_0^3
\end{equation}
 contained within a shell of radius $R_0$, which decreases as 
\begin{equation}\label{MNM0}
  M(N)=e^{-N}M_0 \,.
\end{equation}
At this point, we can envisage two extreme possibilities associated with different choices of the collapsing radius $R_0$. First, we can consider an \textit{early screening} scenario where $R_0$ is identified with the radius of the initial horizon, namely $R_{\rm in}\equiv H^{-1}_{\rm in}$. This corresponds to a situation in which the screening mechanism is highly efficient and only those particles within the Hubble radius at the moment the fifth force starts acting can experience it and end up locked into virialized halos. Second, we can contemplate a \textit{late screening} setup in which all the growing shells within the Hubble radius at virialization, $R_V\equiv H_{\rm V}^{-1}$, falls into the primordial dark matter halo before the fifth force is fully screened. 
It is likely that the actual screening process will take place somewhere within these two limiting cases, which we now discuss in detail:

\begin{enumerate}[label=(\roman*),wide, labelwidth=!]
\item \textit{Early screening}: If we identify the radius $R_0$ in Eq.~\eqref{M0R} with that associated with the initial horizon $R_{\rm in}\equiv H^{-1}_{\rm in}$, only an initial mass
\begin{equation}\label{eq:massradhorcase1}
    M_0=\frac{4\pi}{3}\frac{\rho_\psi(a_{\rm in})}{H_{\rm in}^3} \simeq \frac{4\pi}{3\beta^2}\frac{M_P^2}{H_{\rm in}}
\end{equation}
will collapse into a PDMH. Taking into account the reduction factor \eqref{MNM0} following from the variation of the $M_0$ constituents  up to virialization, we get a PDMH mass
\begin{equation}\label{MNVM0}
M_{\rm PDMH}=e^{-N_{\rm V}}M_0 \,,
\end{equation}
with $M_{\rm PDMH}\equiv M(N_{\rm V})$ and  $N_{\rm V}=\log(a_{\rm V}/a_{\rm in})$. Note that $M_{\rm PDMH}$ is significantly smaller than the mass contained in the horizon  at that time, namely
\begin{equation}
  M_H(N_{\rm V})=\frac{4\pi}{3}\frac{\rho_\psi(N_{\rm V})}{H^3(N_{\rm V})} \simeq \frac{4\pi}{3\beta^2}\frac{M_P^2}{H(N_{\rm V})}\,,
\end{equation}
where in the last step we have employed the value of the density parameter $\Omega_\psi$ according to the scaling solution \eqref{eq:phaseq-1}.  Indeed, taking into account that $M\sim H^{-1}\sim a^{-2}$ we get 
${M_H(N_{\rm V})=e^{2N_{\rm V}} M_0}$ and 
\begin{equation}\label{MNM1}
M_{\rm PDMH}=e^{-3N_{\rm V}}M_H(N_{\rm V}) \,.
\end{equation}
Using  Eqs.~\eqref{aVaeq}, \eqref{eq:massradhorcase1} and \eqref{MNVM0} we can obtain explicit relations among the coupling $\beta$, the initial radius of the fluctuation and the PDMHs mass \cite{Amendola:2017xhl}, namely
\begin{eqnarray}\label{beta585}
\frac{\vert\beta\vert}{585}&=& \, e^{-\frac{N_{\rm V}}{2}} \left(\frac{M_{\rm PDMH}}{M_\odot}\right)^{-1/6}
\,,  \\
H^{-1}_{\rm in}  &\simeq &  2\times 10^{-2}\left(\frac{M_{\rm PDMH}}{M_\odot}\right)^{2/3} {\rm AU}\,, \label{Hin}
\end{eqnarray}
with ${\rm AU}=1.49\times 10^8\,{\rm km}=4.85\times 10^{-12}$ Mpc denoting astronomical units. 

A very rough estimate of the mass-radius relation can be obtained by assuming virialization to occur close to a critical density contrast $\delta_c\sim {\cal O}(1)$. Combining this educated guess with Eqs.~\eqref{eq:Revol} and \eqref{Hin} and assuming ${R_0=R_{\rm in}\equiv H^{-1}_{\rm in}}$ and $\delta_{\psi,\rm in}\ll 1$ we get ${R_{\rm PDMH}\sim H^{-1}_{\rm in} e^{N_{\rm V}}}$ or equivalently
\begin{equation}\label{eq:virradiusapprox}
R_{\rm PDMH}\sim  440 \cdot e^{N_{\rm V}-10}\,  \left(\frac{M_{\rm PDMH}}{M_\odot}\right)^{2/3}  \textrm{AU}\,,
\end{equation}
with 
\begin{equation}\label{eq:onset_nl}
N_{\rm V}= \ln\left(\frac{a_{\rm V}}{a_{\rm in}}\right)=\frac{1}{p}\ln\left(\frac{\delta_c}{\delta_{\psi,{\rm in}}}\right)\,.
\end{equation}
The typical values of $N_V$ are ${\cal O}(10)$. Together with Eq.~\eqref{eq:av3}, this yields an estimate of the value of $a_{\rm in}$ needed for PDMHs to constitute the entire dark matter. For $\beta=4$ one has
\begin{equation}
a_{\rm in}\approx 3\cdot 10^{-6}\,,\hspace{15mm} z_{\rm in}\approx 3\cdot 10^8\,.
\end{equation}
Additionally, combining Eqs. \eqref{beta585} and \eqref{eq:onset_nl}, we can derive an estimate of the coupling to mass relation as a function of the initial density contrast $\delta_{\psi,\rm in}$, namely 
\begin{equation}\label{eq:gen-beta-M}
   \vert\beta\vert\simeq \,585 \left(\frac{\delta_c}{\delta_{\psi,{\rm in}}}\right)^{-1/2p} \left(\frac{M_{\rm PDMH}}{M_\odot}\right)^{-1/6}\,.
\end{equation}
Taking into account the nucleosynthesis constraint $\beta>3$ (cf. Sec. \ref{sec:background}), this relation translates into an upper bound on $M_{\rm PDMH}$, which as shown in Fig.~\ref{fig:bvsd} is very sensitive to $\delta_{\psi,\rm in}$.

 An accurate estimate $\delta_c \simeq 2.07$ (with a weak dependence on $\beta$ and on $\delta_{\psi, \rm in}$) can be obtained by numerically following the evolution of the system according to Eq.~\eqref{eqn:eqnvir}. The result of this procedure is shown in Figs.~\ref{fig:SolnD}, \ref{fig:Radius} and \ref{fig:NL}.  Figures \ref{fig:RvsD} and \ref{fig:bvsd} display the resulting radius and mass of PDMHs as a function of the initial density contrast $\delta_{\psi, \rm in}$. For a fiducial value $\delta_{\psi, \rm in}=10^{-6}$, we obtain a mass-radius relation
\begin{equation}\label{eq:virradius2}
 R_{\rm PDMH}=100 \left(\frac{M_{\rm PDMH}}{M_\odot} \right)^{2/3}  \textrm{AU}\,,
\end{equation}
and a mass bound $M_{\rm PDMH}< 16 M_\odot$ for $\beta>3$. cf.~Eq.~\eqref{eq:gen-beta-M}. Note, however, that this upper limit is very sensitive to the initial density contrast $\delta_{\psi,\rm in}$, as clearly appreciated in Fig.~\ref{fig:bvsd}. Indeed, for an initial value ${\delta_{\psi, \rm in}=5\cdot 10^{-4}}$, we obtain a much less restrictive bound $M_{\rm PDMH}< 10^{4} M_\odot$.

\begin{figure}
\hspace{-0.2cm}\centering
\includegraphics[scale=1.08]{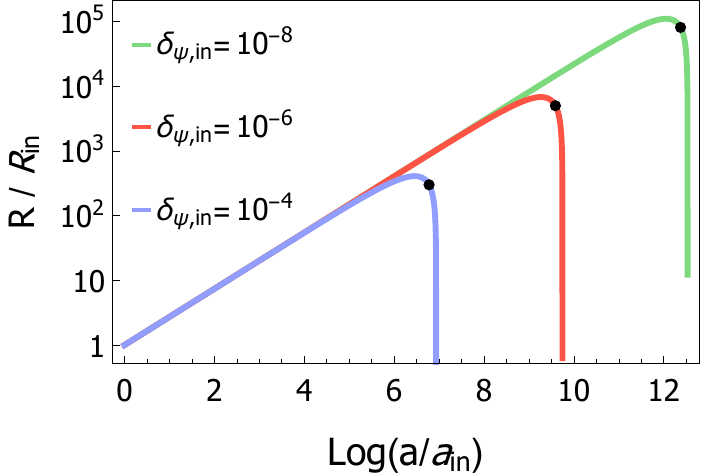}
\caption{Evolution of the overdensity radius as a function of the number of e-folds $\log a/a_{\rm in}$, with $R_{\rm in}\equiv H_{\rm in}^{-1}$  the initial horizon radius. The curves reach their maximum at turnaround and the black dots indicate the radius at which the virialization condition \eqref{eqn:eqnvir} becomes satisfied.}\label{fig:Radius}
\end{figure}
\begin{figure}
\hspace{0cm}\centering
\includegraphics[scale=0.79]{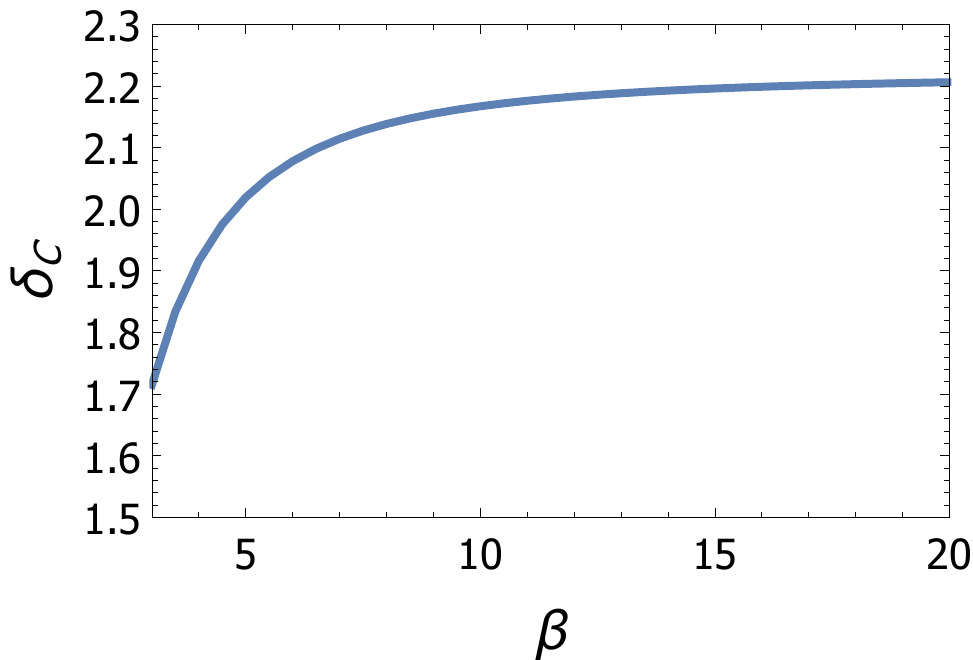}
\caption{Critical density contrast $\delta_c$ as a function of $\beta$ for an initial density contrast $\delta_{\rm \psi,in}=10^{-6}$. For sufficiently large couplings this quantity saturates to $\delta_c\simeq 2.2$. This trend turns out to be independent from the initial condition on $\delta_{\rm \psi,in}$.}\label{fig:NL}
\end{figure}
\begin{figure}
    \hspace{-0.3cm}\centering
    \includegraphics[scale=0.8]{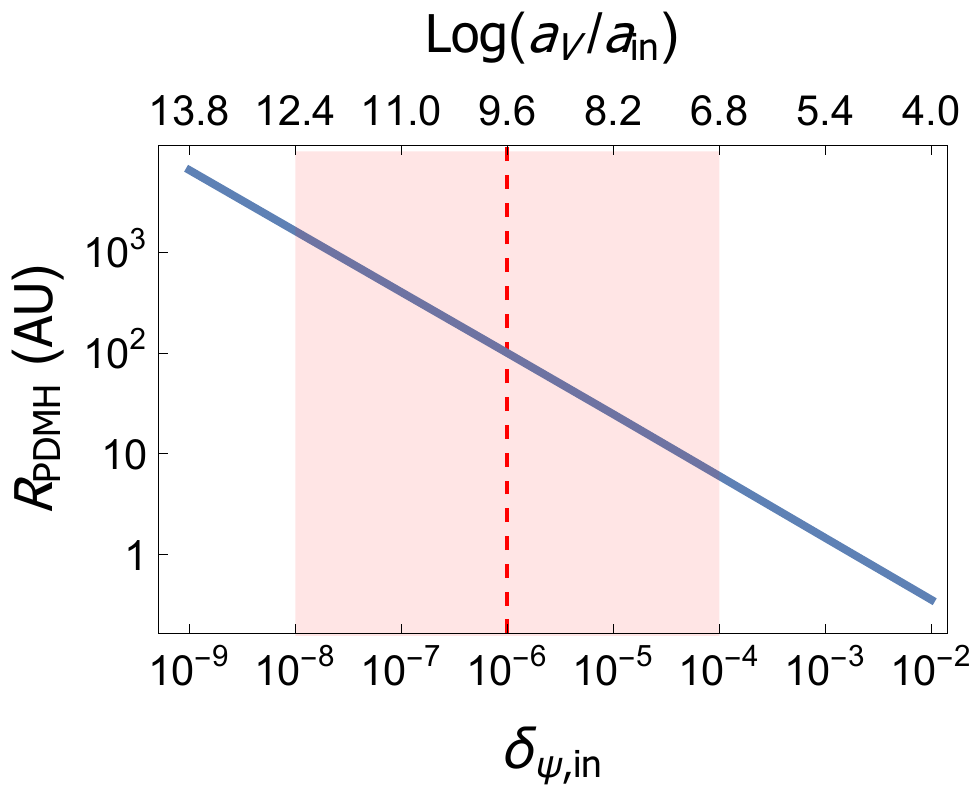}
    \caption{Solar-mass PDMHs radius following from the numerical evolution of the system according to Eq.~\eqref{eqn:eqnvir} as a function of the number of e-folds $N_V=\log(a_V/a_{\rm in})$ from the onset of the scaling regime to virialization. As illustrated in the figure, the value of $N_V$ depends on the initial density contrast at solar mass scales, which we assume to be in the range $10^{-8} \leq\delta_{\psi,{\rm in}}\leq 10^{-4}$ (cf. Sec. \ref{sec:growth}). In accordance with Eqs.~\eqref{eq:virradiusapprox} and \eqref{eq:onset_nl}, the slope of the blue line is proportional to $-1/p$.}
    \label{fig:RvsD}
\end{figure}

\item \textit{Late screening}:
If the matter surrounding the growing perturbation within the initial horizon radius $H_{\rm in}^{-1}$ fall into the primordial dark matter halos before the fifth force is completely screened, the above estimates should be modified. To evaluate the impact of this potential infall, we focus on the limiting situation in which the whole dark matter component within the horizon radius at virialization is locked into a halo.  In this case, we get a much more compact PDMH with radius 
\begin{equation}\label{eq:virradius3}
R_{\rm PDMH}\equiv H_{\rm V}^{-1}=2\times 10^{-2} \left(\frac{M_{\rm PDMH}}{M_\odot} \right)^{2/3} \textrm{AU}\,.
\end{equation}
\end{enumerate}

\begin{figure}
    \centering
    \includegraphics[scale=0.90]{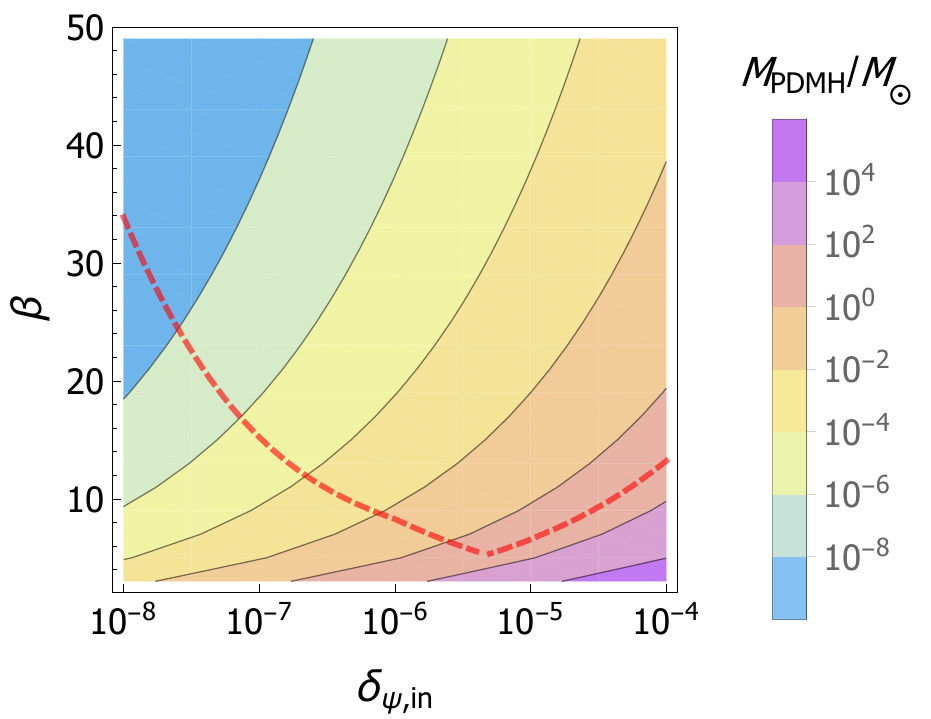}
    \caption{Mass of the PDMHs as a function of the initial density contrast $\delta_{\rm \psi,in}$, and the fifth force coupling, $\beta$. The region below the red line represents the set of parameters evading microlensing constraints; cf.~Sec. \ref{sec:micro}. The ending upward turn reflects the upper limit of the microlensing constraints mass window extension.}
    \label{fig:bvsd}
\end{figure}

\subsection{Screening}\label{sec:screening}

The treatment leading to the radius estimates in Eqs.~\eqref{eq:virradius2} and \eqref{eq:virradius3} implicitly assumes that, even if the fifth force becomes eventually suppressed outside PDMHs, it remains active within them. According to Refs.~\cite{Casas:2016duf,Bonometto:2017lhg,Bonometto:2017irg}, this could make the virialization stage transitory and lead to the eventual dissolution of the halos. Note, however, that there are many ways in which this could be avoided. One could consider for instance screening scenarios in which --in clear analogy with electrostatics-- the scalar charge in a PDMH would end up confined to a very thin shell near its surface \cite{Khoury:2003aq,Khoury:2003rn}.
Alternatively one could envisage a multifermion dynamics with a locking mechanism \cite{Farrar:2003uw,Gubser:2004du} or a potential relaxation of the constituent masses to a constant value compatible with the above dynamics \cite{Bonometto:2018dmx}. 

Whatever the mechanism stopping the evolution of the constituent masses, the effective coupling $Y$ in Eq.~\eqref{eqn:eqnvir} would approach unity in the latest stages of PDMHs formation, leading to an increase of the virialization radius for a given mass by a factor $\sim 3.5$ as compared to the estimates in Sec. \ref{sec:massradius}.
As will become clear in Sec. \ref{sec:CMB} and \ref{sec:micro}, this would not affect our conclusions regarding compatibility with data, but rather strengthen them.
For this reason we will stick to the conservative values \eqref{eq:virradius2} and \eqref{eq:virradius3} in what follows.

\subsection{Mass of the $\psi$-particles}\label{sec:masses}
\begin{figure}
    \centering
    \hspace{-0.3cm}\includegraphics[scale=1.1]{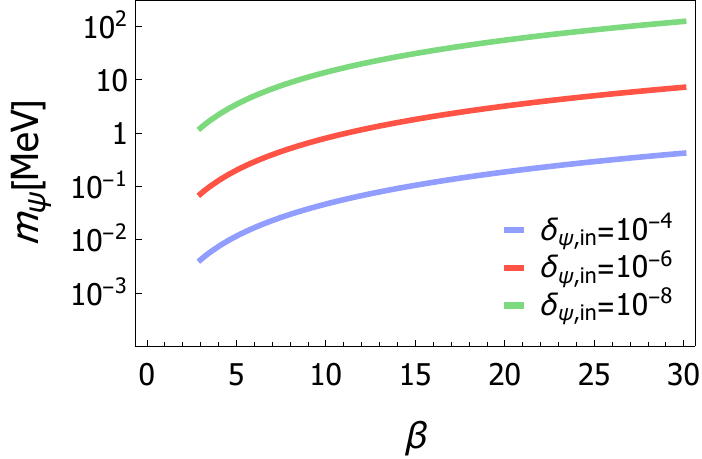}
    \caption{Initial $\psi$-particle mass $m_{\psi}(a_{\rm in})$ as a function of $\beta$ for various choices of the initial density contrast $\delta_{\psi,{\rm in}}$.}
    \label{fig:psimass}
\end{figure}

An order of magnitude estimate for the bare mass parameter to be inserted  into Eq.~\eqref{eq:action3}  can be obtained by considering the temperature scales involved in PDMHs formation. As a first guess, we assume  thermal equilibrium. Using the standard relation\footnote{We ignore again an order unity correction ${\gamma(a)\equiv(g_{\rm in}/g_{\rm eq})^{-1/3}}$, with $g_{\rm in}$ the initial number of relativistic degrees of freedom and $g_{\rm eq}\approx 3.36$ its value at matter-radiation equality.} $T\sim a^{-1}$ together with Eqs. \eqref{aVaeq} and \eqref{eq:massin} and omitting order one factors, we get 
\begin{equation}
T(a_{\rm in})=T(a_{\rm V}) e^{N_{\rm V}}\approx \beta^2\,T_{\rm eq}e^{N_{\rm V}}
\end{equation}
for the temperature at the onset of the scaling solution.
In order for the $\psi$ particles to be non relativistic at this temperature, and therefore to feel the fifth force, their masses must exceed $T(a_{\rm in})$. For $m(a_{\rm in})=T(a_{\rm in})$ the fermion mass at
virialization must be of the order of $m_\psi(a_{\rm V})\approx $\, 0.01 keV, 0.1 keV, 1 keV for $T_{\rm eq}\simeq {\cal O}({\rm eV})$, $N_{\rm V}\simeq 10$ and $\beta= 3,\, 10,\, 30 $, respectively.  This corresponds to masses  $m_\psi(a_{\rm in})\approx$\, 0.1 MeV, 1 MeV, 10 MeV at the onset of the scaling regime, meaning that this occurs just around the epoch of primordial nucleosynthesis. In Fig.~\ref{fig:psimass} we plot the initial fermion mass as a function of $\beta$ for various $\delta_{\psi,{\rm in}}$.
Dropping the assumption of thermal equilibrium, the momentum distribution of the $\psi$-particles may still be peaked at the radiation temperature. The condition for the fermions to be non relativistic becomes then $m(a_{\rm in})\ge T(a_{\rm in})$. Our computed values should be then understood as lower bounds. In fact, higher masses that become non relativistic earlier might still reach the scaling solution at the same time $a_{\rm in}$ if their initial density is sufficiently low. In this case, they would pass through an intermediate regime in which their energy density decays slower than radiation before finally joining the scaling attractor (\ref{eq:phaseq-1}) at $a_{\rm in}$.

\section{Observational constraints} \label{sec:constraints}

In the absence of decays or annihilations of the constituent dark matter particles, PDMHs  will not be restricted by gamma-ray observations \cite{Scott:2009tu,Ackermann:2015zua}. Potential constraints on these objects could come, however, from (i)~CMB energy injection bounds \cite{Ali-Haimoud:2016mbv} (ii) microlensing observations \cite{Alcock:1993qc,Aubourg:1993wb,Udalski:1993zz,Auriere:2001ha,Riffeser:2003rs,Wyrzykowski:2010mh,Wyrzykowski:2011tr}, (iii) type Ia supernovae datasets~\cite{Zumalacarregui:2017qqd}. 

For halo masses in the solar mass range ${M_{\rm PDMH}\sim M_\odot}$, the PDMHs radius following from an early screening and a fiducial initial density contrast, $\delta_{\rm \psi,in}\simeq 10^{-6}$, is roughly 3-4 times the distance to Neptune and therefore much larger than the Schwarzschild radius ${R_S\sim (M_{\rm PDMH}/M_\odot)}$ km in the same solar mass range. On top of that, the trend $(M_{\rm PDMH}/M_\odot)^{2/3}$ in Eq.~\eqref{eq:virradius2} implies that $R_{\rm PDMH}$ grows faster with $M_{\rm PDMH}$ than the Einstein radius $R_E\sim (M_{\rm PDMH}/M_\odot)^{1/2}$ and therefore that, for a sufficiently large mass, the halo will be larger than $R_E$. In the following sections we show that these two properties are indeed enough to pass the observational constraints i) and ii). The more involved analysis of supernovae datasets is left for future work.

\subsection{Cosmic Microwave Background constraints}\label{sec:CMB}

The radiation emitted during matter infall into collapsed structures modifies the reionization history. The CMB injections bounds due to primordial black holes have been recently reassessed in the literature. In particular, the results of Ref.~\cite{Ali-Haimoud:2016mbv} show that the consistency of both the temperature and polarization spectra forbids these objects to account for the total dark matter component if their masses are in the range ${10^2 M_\odot<M<10^4 M_\odot}$. Note, however, that the luminosity of primordial black holes is mostly due to the Bremsstrahlung radiation emitted in the vicinity of the Schwarzschild radius since it is there where the accreted gas acquires relativistic velocities. Since our PDMHs are significantly larger than their Schwarzschild radius, we can foresee that the CMB constraints on them are doomed to disappear. 

To make the above statement quantitative, we  follow closely the analysis in Ref.~\cite{Ali-Haimoud:2016mbv}. In particular, we  consider the radial accretion of hydrogen onto an isolated PDMH of mass  
$M$ surrounded by the almost uniform CMB radiation fluid.  The integrated luminosity of the fully ionized thermal electron-proton plasma is given by 
\begin{equation}
L=4\pi\int j\,r^{2}dr
\end{equation}
with $r$ a radial coordinate and
\begin{equation}
j=\alpha\, \sigma_{T}\, n_{e}^{2}\, F(T)\,
\end{equation}
the frequency-integrated emissivity. Here $\alpha$ denotes the fine-structure constant, $\sigma_T$ is the Thomson cross section and $n_{e}$ stands for the electron number density
\begin{equation}\label{eq:ne}
n_{e}=\frac{\dot{M}}{4\pi m_{p}r^{2}|v|}
\end{equation}
with $m_{p}$ the proton mass,
\begin{equation}\label{eq:vinfall}
|v|=\sqrt{\frac{R_{S}}{r}}\,
\end{equation}
the infall velocity at a distance $r$ and $R_S=2 \,G M$ the Schwarzschild radius.
The quantity 
\begin{equation}\label{Fdef}
F(T)\equiv T\, J\left(X\right)\,,    
\end{equation}
with
\begin{equation}
J(X) \simeq
\begin{cases}
\frac{4}{\pi} \sqrt{\frac{2}{\pi}} X^{-1/2} \left(1 + 5.5 X^{1.25} \right),  & X < 1\\
\frac{27}{2 \pi} \left[\ln(2 X e^{- \gamma_{\rm E}} + 0.08) + \frac{4}{3}\right],  & X > 1~~
\end{cases}
\end{equation}
a dimensionless function of the temperature $T$ over the electron mass $m_e$ and $\gamma_{\rm E} \approx 0.577$ the Euler's constant, scales with the temperature as $F(T)\sim r^{-1}$ \cite{Ali-Haimoud:2016mbv,Svensson:1982hz}. Expressing this function in terms of its value at the boundary of the emitting sphere,
$ F(T)=F(T_R)/r$, and taking into account Eqs.~\eqref{eq:ne} and \eqref{eq:vinfall}, we can express the radiative efficiency  $\varepsilon\equiv L/\dot{M}$ as
\begin{equation}\label{eq:Reff}
\varepsilon =\frac{\alpha}{2m_{p}}\frac{\dot{M}}{L_{\rm Edd}}\frac{F(T_{R})}{R}\,,
\end{equation}
with 
\begin{equation}
L_{\rm Edd}= \frac{2 R_S m_{p}}{\sigma_{T}}
\end{equation}
the Eddington luminosity.\footnote{This is defined as the maximum luminosity of a source in hydrostatic equilibrium.} This expression coincides with the primordial black hole radiative efficiency computed in Ref.~\cite{Ali-Haimoud:2016mbv} when the emitting boundary is identified with the Schwarzschild $R=R_S$ and the function $F$ is appropriately rescaled as $F(T_R)=F(T_S) R_s$. Note, however, that the PDMH radii computed in the previous section are generically much larger than $R_S$. This translates into a substantial reduction of the radiative efficiency in Eq.~\eqref{eq:Reff} as compared to the primordial black hole case. The analysis presented in Appendix \ref{sec:tempboundary} yields 
\begin{equation}
\frac{F(T_{R})}{F(T_{S})}  \ll1\,.
\end{equation}
Since the energy deposit is proportional to $F(T)$, we can conclude that this is significantly smaller
for PDMHs as compared to primordial black holes.

\subsection{Microlensing constraints}\label{sec:micro}

The amount of primordial black holes playing the role of dark matter in the mass window from $10^{-8}$ to $10\,M_{\odot}$ is strongly constrained by microlensing observations  \cite{Alcock:1993qc,Aubourg:1993wb,Udalski:1993zz,Auriere:2001ha,Riffeser:2003rs,Wyrzykowski:2010mh,Wyrzykowski:2011tr}. 
Pointlike objects with a mass larger than $10\,M_{\odot}$ produce microlensing patterns on timescales larger than the typical observation times of MACHO and EROS collaborations, and therefore microlensing constraints do not extend above this mass.
If the radius of the dark matter halos is smaller than the Einstein radius, they act essentially as pointlike lenses and the stringent microlensing constraints on primordial black holes inevitably apply to them, as to similar \textit{compact} objects.  However, the pointlike approximation breaks down for sufficiently large PDMHs,  which should then be described as extended lenses, as we do below. As compared to a pointlike lens with the same mass, an extended lens takes longer to provide a complete microlensing pattern. Therefore, the heaviest PDMHs that the examined microlensing experiments are able to constrain could be potentially lighter than $10\,M_{\odot}$, being this a rather conservative value. 

In the case of PDMHs with a radius larger than the Einstein radius, an estimate of this bound can be obtained approximating\footnote{This value should be understood just as an order of magnitude estimate. In particular, since the distance where the entire event takes place is clearly larger than $R_{\rm PDMH}$  we should generically expect a lower mass threshold.} the timescale of microlensing phenomena as ${T=R_{\rm PDMH}/v}$, with ${v\simeq200\,\textrm{km/s}}$ \cite{Paczynski:1985jf}. The longest period of microlensing data acquisition of the MACHO and EROS collaborations is about $ 6 \,\textrm{yr}$ \cite{Alcock:2000kd,PalanqueDelabrouille:1997uj}. Combining Eqs.~\eqref{eq:virradiusapprox} and \eqref{eq:onset_nl}, the heaviest PDMHs mass to whom microlensing constraints can extend, $M_{\rm PDMH}^{*}$, is then related to $\delta_{\rm \psi,in}$ as
\begin{equation}\label{eq:uppmass}
    M_{\rm PDMH}^{*}\simeq 1.4\cdot 10^{6}\,(\delta_{\rm \psi,in})^{1/p}\,M_{\odot}\,.
\end{equation}
Adopting a fiducial initial value $\delta_{\rm \psi,in}\simeq 10^{-6}$, PDMHs become constrained by microlensing experiments up to $M_{\rm PDMH}^{*}\simeq 3.36\,M_\odot$. In general, the higher the initial density contrast, the lower this bound.

The Einstein radius for MACHO/EROS microlensing phenomena is given by \cite{1994ApJ...430..505W} 
\begin{equation}
    R_E\simeq 21.1 \left(\frac{M}{M_\odot}\right)^{1/2}\left[\xi(1-\xi)\right]^{1/2}\,\mathrm{AU}\,,
\end{equation}
with $\xi=w_\mathrm{d}/w_\mathrm{s}$, $w_\mathrm{d}$ ($w_\mathrm{s}$) the distance between the observer and the deflector (source) and $M$ the mass of the lens, identified in our case with that of the PDMHs. The parameter $\xi$ is restricted to the range $0<\xi<1$ with $\xi=1/2$ corresponding to a lens equally distant from the source and the observer. In this case the Einstein radius is maximized and amounts to $R_{E}\simeq 10.7 \, (M/M_{\odot})^{1/2}\, \mathrm{AU}$. 

If we look at the solar and subsolar mass window, where the microlensing constraints are effective, the radius of the PDMHs in the {\it late screening} scenario is always smaller than the Einstein radius. Therefore, they are regarded as pointlike lenses and ruled out from providing the whole dark matter component within this mass range.

In the {\it early screening} case, the virialization radius in Eq.~\eqref{eq:virradius2} for a fiducial density contrast $\delta_{\rm \psi,in}=10^{-6}$, is bigger than the Einstein radius for a halo masses  $M_{\rm PDMH}>10^{-6}M_\odot$. Within this mass range, the primordial black hole constraints do not directly apply to PDMHs and must be reconsidered using an extended lens configuration, as we illustrate below. 

\subsubsection{Cored isothermal sphere}

To study how microlensing constraints are modified when the predicted size of our structures is taken into account, it is necessary to make some assumption on the PDMHs density profile. The formation process of this density profile is, however, complicated to model due to the highly nonlinear character of the problem at hand and the strong coupling regime under consideration. As a matter of fact, even the characterization of the inner density profile of dark matter halos in a $\Lambda$CDM cosmology is still quite debated. While \textit{N-body} simulations within the concordance model predict the appearance of cuspy profiles both in early- \cite{Gosenca:2017ybi,Delos:2019mxl,Delos:2017thv} and late-time \cite{Navarro:1996gj,Diemand:2005vz} clumped dark matter structures, different observations of dwarf spheroidals  \cite{Swaters:2002rx} and low surface-brightness disk galaxies \cite{deBlok:2001hbg,deBlok:2005qh,Oh:2008ww} do \textit{not} agree with this prediction, but rather suggest the existence of central cores at small radii (for a review, see Ref.~\cite{deBlok:2009sp}). 

It is also worthwhile to notice that the appearance of cuspy density profiles in dark matter structures \cite{Navarro:1996gj,Diemand:2005vz,Gosenca:2017ybi,Delos:2019mxl,Delos:2017thv} is strictly associated to $\Lambda$CDM and does not necessarily apply to other cosmological scenarios. In particular, the above numerical simulations are intrinsically Newtonian and cannot account for additional interactions unless suitably modified.

In the lack of a proper non-Newtonian $N$-body simulation able to account for fifth force effects during radiation domination, we will assume our PDMHs to be similar to the observed structures in the Universe, i.e. noncuspy. For illustration purposes, we will assume their density distribution to be described by a nonsingular isothermal  profile\footnote{For simplicity,
we assumed the core of the density distribution to coincide with PDMHs virialization radius. A more general definition of the core radius is $R_{c}=R_{\rm PDMHs}/c$, with $c$ a constant parameter. Picking $c\simeq10$ would slightly change the analysis presented in this section as it can be compensated by employing a smaller fiducial value $\delta_{\rm \psi, in}\sim 10^{-7}$. } \cite{1987gady.book.....B}
\begin{equation}\label{eq:profile}
    \rho=\frac{\rho_0}{1+(r/R_{\rm PDMH})^2}\,,
\end{equation}
with $r$ the radial distance from the center of the sphere and $\rho_0$ the density at the center of the mass distribution, obtained normalizing the mass enclosed by the virialization radius $R_{\rm PDMH}$ in Eq.~\eqref{eq:virradius2} to the mass of the halo.
This very common profile in the literature of dark matter halos describes a system of collisionless particles in hydrostatic equilibrium \cite{Salucci:2018hqu}.
Note, however, that other choices describing a cored system are possible as well. For instance, one could consider a Burkert profile \cite{Burkert:1995yz} without significantly altering the results below, as we have explicitly verified. In this sense, our conclusions can be considered independent from the profile choice in Eq.~\eqref{eq:profile}. 

\begin{figure}
\hspace{-0.3cm}\centering
\includegraphics[scale=0.8]{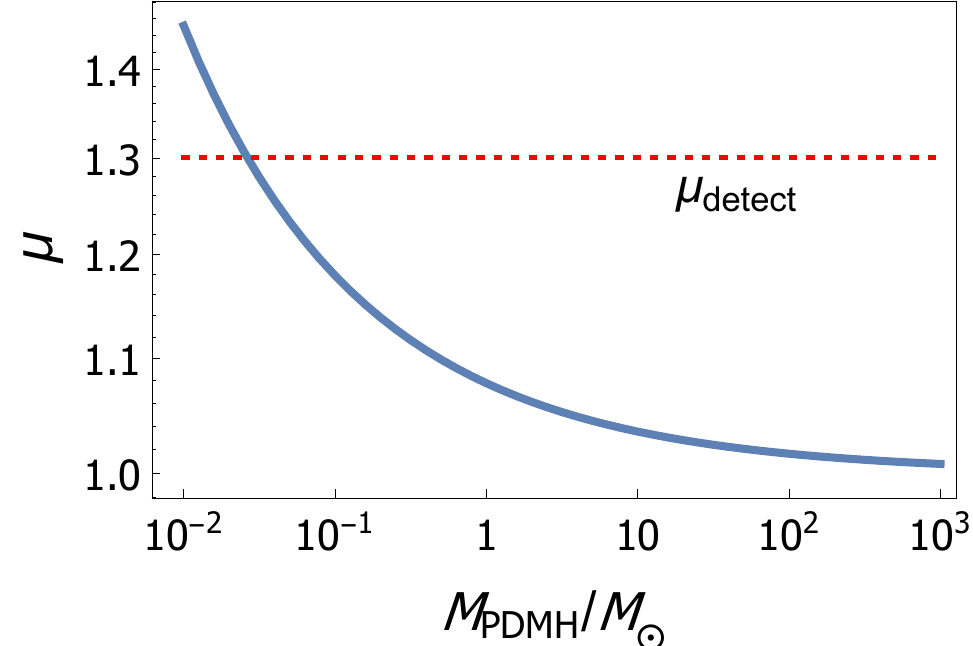}
\caption{\textit{Early screening scenario.} Maximum magnification $\mu$ generated by PDMHs as a function of their mass for a fiducial density contrast $ {\delta_{\rm \psi,in}=10^{-6}}$, with a lens configuration characterized by ${\xi=1/2}$. The red dotted line displays the MACHO
collaboration  identification  threshold.}
\label{fig:Magn}
\end{figure}

\begin{figure}
\hspace{-0.3cm}\centering
\includegraphics[scale=1.25]{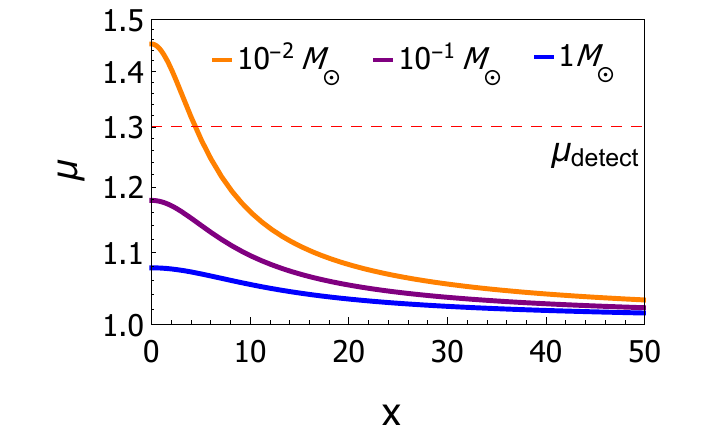}
\caption{\textit{Early screening scenario.} Magnification curves generated by PDMHs as a function of the observed image position $x$ for a fiducial density contrast $\delta_{\rm \psi,in}=10^{-6}$.}\label{fig:Magn2}.
\end{figure}

\begin{figure}
    \hspace{-0.2cm}\centering
    \includegraphics[scale=0.8]{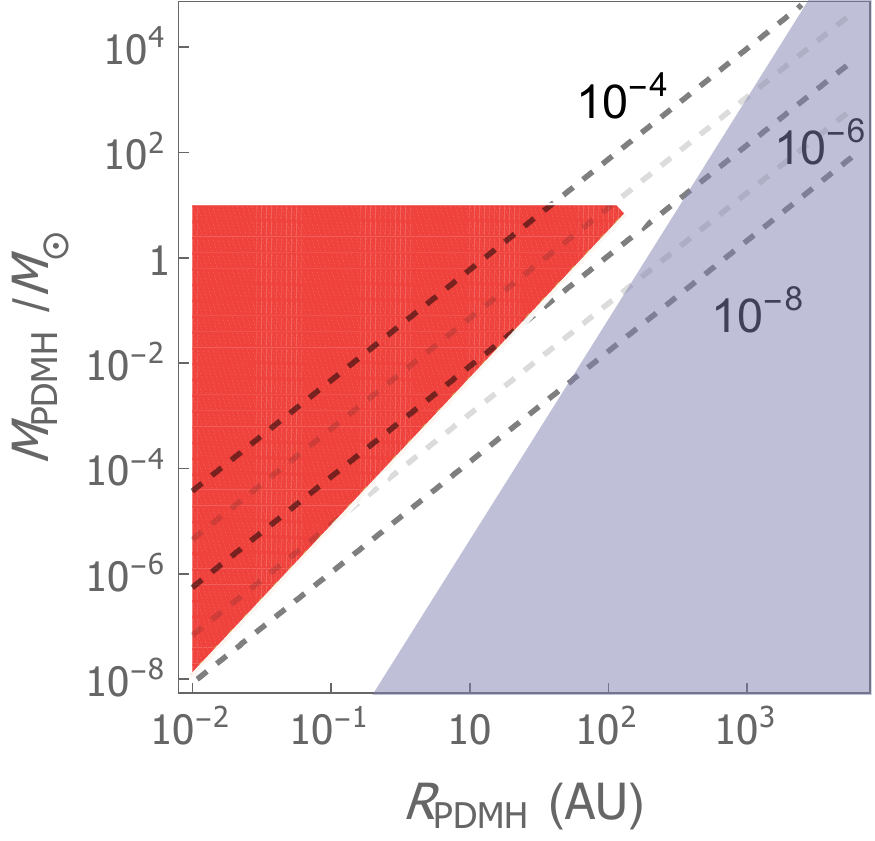}
    \caption{\textit{Early screening scenario.} The white region represents the allowed parameter space.  The red region displays the set of physical configurations ruled out by microlensing experiments.
    The blue region identifies the  values ($M_{\rm PDMH},R_{\rm PDMH}$) precluded by nucleosynthesis  requirements (i.e. with $\beta>3$). 
       The black dashed lines track the mass-radius relation resulting from our model for distinct values of the initial density contrast $\delta_{\rm \psi,in}$.}
    \label{fig:magcont}
\end{figure}

For masses in the solar range, the virialization radius is significantly smaller than the distance to standard microlensing sources, which include, among others,  stars in the Large Magellanic Cloud at around $50$ $ \textrm{Kpc}$. This hierarchy allows us to describe the PDMHs as thin lenses with respect to the line of sight. We can additionally benefit from the axial symmetry of the problem to describe the lensing phenomenon in terms of a single deflection angle $\alpha$ on the plane spanned by the positions of the source, the observer and the lens. Within this setup, the lens equation reads
\begin{eqnarray}\label{lenseqn}
y(x)&=&x-\alpha(x)=x-\frac{x_0}{x}\left[L(x)-r_{\rm PDMH}\right]\,,
\end{eqnarray}
with $y,x$ standing respectively for the real and observed position of the source on the deflector plane rescaled to the Einstein radius,  $r_{\rm PDMH}\equiv R_{\rm PDMH}/R_E$, and \
\begin{equation}
L^2(x)\equiv x^2 +r_{\rm PDMH}^2   \,, \hspace{7mm} x_0\equiv \frac{2R_E\pi}{(4-\pi)R_{\rm PDMH}}\,.
\end{equation}
Since Eq.~\eqref{lenseqn} has a unique solution for
\begin{equation}
    r_{\rm PDMH}>2\,,
\end{equation}
the lensing generated by each PDMH results in a single deflected image for $M>10^{-5} M_\odot$. The magnification $\mu$ is defined as the inverse of the lens mapping determinant and can be recast as
\begin{equation}
    \mu^{-1}(x)=\left(1-\frac{x_0}{2 L(x)}\right)^2   -\frac{x_0^2\left(L(x)-r_{\rm PDMH}\right)^4}{4x^4L^2(x)}\,.
\end{equation}
In the single image regime, there is only a contribution that adds up to the total magnification function. For our purposes, it would be enough to consider just the maximum value this function can reach, regardless of the position of the source or the image at which this is achieved. An explicit solution of Eq.~\eqref{lenseqn} is therefore not required. The analysis of the total magnification function is carried out in Figs.~\ref{fig:Magn} and \ref{fig:Magn2} for a fiducial density contrast $\delta_{\rm \psi,in}=10^{-6}$.
Comparing these plots with the  identification threshold ${\mu>\mu_{\rm detect}\simeq 1.30}$ of the MACHO and EROS collaborations \cite{Alcock:2000kd,1998astro.ph.12173E}, we can identify a  mass window $M_{\rm PDMH}\gtrsim 0.03 M_\odot$ where the PDMHs cannot be detected by current  microlensing experiments.
This result, combined with the condition $M_{\rm PDMH}< 16 M_\odot$ inferred from nucleosynthesis, identifies a viable mass window from $0.03$ to $16\, M_\odot$.
Note, again, that this range strongly depends on the initial density contrast $\delta_{\rm \psi,in}$, as explicitly shown in Fig.~\ref{fig:magcont}. The window where PDMHs are compatible with microlensing observations  extends from $10^{-8}$ to $10^{4}\,M_{\odot}$ for an initial density contrast $\delta_{\rm \psi,in}$ within $10^{-4}/10^{-8}$.

\section{Conclusions}\label{sec:conclusions}

We argued that primordial dark matter halos could be generated at very large redshifts within the radiation dominated era. The necessary ingredients are i) a light scalar field $\phi$ mediating an attractive interaction stronger than gravity and ii) some heavy degrees of freedom $\psi$ strongly interacting with it with a suitable abundance. While the light scalar field could be potentially identified with a dark energy component, the heavy degrees of freedom could play the role of usual dark matter candidates. If the interaction among these two species is large enough, the system enters a scaling regime during radiation domination where the primordial perturbations of the heavy field become significantly enhanced. Assuming the eventual screening of the scalar force, we determined the properties of the collapsing matter at virialization.
For an early screening and a fiducial density contrast $\delta_{\psi, \rm in}\sim 10^{-6}$, the PDMHs radius turns out to be significantly larger than the corresponding Schwarzschild and Einstein radii for halo masses between $0.03$ to $16 M_\odot$, being the upper bound imposed by nucleosynthesis constraints.
This makes the created objects unobservable by current microlensing experiments and significantly reduces their energy injection in the Cosmic Microwave background as compared to primordial black holes. After due consideration of the various constraints, we find that a successful scenario requires an effective coupling $3\lesssim \beta \lesssim 30$ and initial $\psi$-particle masses larger than $0.1$ MeV. 
This translates into a \textit{formation} redshift $z\sim {\cal O}(10^4-10^6)$, significantly exceeding the one associated with the first DM clumps in a $\Lambda$CDM scenario. 
The initial PDMHs distribution is essentially monochromatic, with a peak mass that, after accounting for the uncertainties on the initial density contrast $\delta_{\psi, \rm in}$, lies between $10^{-8}$ and $10^4$ solar masses. For all practical purposes, the created objects behave just like macroscopic dark matter ``particles" ranging in size from $10^{-2}$ to $10^3$ AU and having an average density 
\begin{eqnarray}
\bar \rho_{\rm PDMH} &\simeq& \,2.2\times 10^{-13}\left(\frac{M_\odot}{M_{\rm PDMH}}\right) \frac{\rm gr}{{\rm cm}^{3}}\,, 
\end{eqnarray}
and an abundance\footnote{We neglect here potential accretion, merging and disruption effects and assume the primordial halos to be distributed in galaxy halos 200 times denser than the cosmological background. i.e. $n_{\rm PDMH} \approx 200\,\rho_{\rm DM}/ M_{\rm PDMH}$. The resulting value of $n_{\rm PDMH}$ is commensurable to the $\Lambda$CDM abundance of dark matter subhalos with masses $>0.01 M_\odot$. Note, however, that a direct comparison of these two values seems hardly feasible given the monochromatic character of the PDMHs distribution as opposed to the power-law distribution of standard subhalos \cite{Springel:2008cc}.}
\begin{eqnarray}
n_{\rm PDMH}&\simeq &  5 \times 10^{13}\, (\Omega_{\rm DM} h^2) \frac{M_\odot}{M_{\rm PDMH}}  \,{\rm Mpc}^{-3}\,.
\end{eqnarray}
Note that although comparable in size, the PDMHs are much denser than the smallest virialized clumps appearing in a $\Lambda$CDM scenario, $\bar \rho^{\rm min}_{ \rm clump}\simeq \,7\times 10^{-22}\, {\rm gr}/{\rm cm}^{3}$ \cite{Berezinsky:2007qu}.
We expect therefore our objects to be more resistant to tidal disruption than standard DM halos.

Even though the results presented in this paper should be understood just as an order of magnitude estimates, the existence of PDMH in the solar mass range constituting the whole dark matter component seems \textit{a priori} plausible within the present uncertainties. Many other interesting aspects such as the precise implementation of the screening mechanism \cite{Zanzi:2015cch} or the resistance to tidal disruptions \cite{Berezinsky:2003vn,Berezinsky:2005py,Berezinsky:2006qm,Berezinsky:2007qu,Kashiyama:2018gsh} are worthy to explore. Among other effects, the survival of PDMHs till the present cosmological epoch could have important consequences for direct dark matter searches. In particular, even if our dark matter particles could be potentially produced at accelerator experiments, they would be hardly observable by direct detection probes due to the drastic reduction of their free number density. 

\section*{Acknowledgments} 

We acknowledge support from the DFG through the project TRR33 ``The Dark Universe'' during the first stages of this work. LA and JR thank S.~Bonometto for discussions. \\

\appendix 

\section{Effective temperature suppression at the boundary}\label{sec:tempboundary}

To determine the factor $F(T_R)$ entering into the energy deposit \eqref{eq:Reff} we benefit from the detailed analysis in Ref.~\cite{Ali-Haimoud:2016mbv}. According to this work, 
\begin{equation}
\hspace{25mm}T_R \simeq  m_{e}\, {\cal F}(Y(R))\,,\hspace{21mm}
\end{equation}
with 
\begin{equation}
{\cal F}(Y)\equiv Y \left(1+\frac{Y}{0.27}\right)^{-1/3}\,,\hspace{5mm}
Y\simeq \gamma\, Y_S\,, 
\end{equation}
and $\gamma\equiv R_{\rm S}/R$. Assuming $X\ll1$ in Eq.~\eqref{Fdef}, together with $Y\gg1$ and $\gamma\ll 1$, we can approximate
\begin{equation}
F(T_{R})\simeq m_{e}\gamma^{1/2}Y_{S}^{1/2}\,,\hspace{10mm}
 T_{S} \simeq 0.65\,m_{e}Y_{S}^{2/3}\,.
\end{equation}
Combining these equations we get 
\begin{equation}
F(T_{R}) \approx1.38\,m_{e}\gamma^{1/2}\left(\frac{T_{S}}{m_{e}}\right)^{3/4}
\end{equation}
and 
\begin{equation}
\frac{F(T_{R})}{F(T_{S})}  =0.01\,\gamma^{1/2}\frac{m_{e}}{m_{p}}\left(\frac{T_{S}}{m_{e}}\right)^{3/4}\left(\frac{\varepsilon}{\dot{m}}\right)^{-1}\,,
\end{equation}
where
$T_{S}/m_e$ and $\varepsilon/\dot{m}$ can be taken from Figs.~5 and 6 in  Ref.~\cite{Ali-Haimoud:2016mbv}. For  $T_{S}/m_{e}\simeq 10^{9}$ and  $\varepsilon/\dot{m}=10^{-5}$ we get a ratio 
\begin{equation}
\frac{F(T_{R})}{F(T_{S})} \simeq \gamma^{1/2}\,,
\end{equation}
which is numerically very small for PDMHs radii much larger than $R_S$. 

\newpage 
\bibliographystyle{apsrev}
\bibliography{References}

\end{document}